# Microporous and Mesoporous Materials
## Amino-Modified ZIF-8 for Enhanced CO2 Capture: Synthesis, Characterization and Performance Evaluation

**ARTICLE**

## Amine-Modified ZIF-2 Capture: Synthesis, Characterization and Performance Evaluation

Viktorie Neubertová,*[a] Václav Švorčík [b] and Zdeňka Kolská [a]

The urgent need for sustainable and innovative approaches to mitigate the increasing levels of atmospheric $CO_2$ necessitates the development of efficient methods for its removal. In this study, we focus on the synthesis and functionalization of metal-organic framework (MOF) ZIF-8 at room temperature to enhance its capacity for $CO_2$ capture. Specifically, we investigated the impact of four amino-compounds, namely tetraethylenepentamine (TEPA), hexadecylamine (HDA), ethanolamine (ELA), and cyclopropylamine (CPA), on the chemical structure, size, surface area and porosity and $CO_2$ capturing of ZIF-8 powder. By varying concentrations of the amino-compounds, we examined their influence on the ZIF-8 properties. These results highlight the potential of simple synthesis and functionalization techniques for MOFs in enhancing their $CO_2$ capture capabilities. The findings from this study offer new opportunities for the development of strategies to mitigate $CO_2$ emissions using MOFs.

## 1. Introduction

In recent decades, there has been growing concern about the excessive accumulation of carbon dioxide ($CO_2$) in the atmosphere and its profound impact on the global environment, as this greenhouse gas is one of the contributors to global climate change[1,2]. Dealing with the ever-increasing amount of $CO_2$ requires innovative and sustainable strategies to mitigate $CO_2$ emissions[3–6]. In this context, there are prominent materials such as activated carbon[7,8], ionic liquids[9–11], mixed matrix membranes (MMMs)[12–14], zeolites[15–17], metal organic frameworks (MOFs)[18–21], and others[22–24], that have gained considerable attention as very promising $CO_2$ sorbents. These materials are known for their characteristic properties and have been intensively investigated for their potential for $CO_2$ capture. The ZIF-8 has attracted considerable attention for a wide range of applications including drug delivery[25,26], catalysis[27], energy storage[28], gas sensing[29], gas separation[30] and storage[31]. ZIF-8 demonstrates a precisely defined porous structure comprised of metal ions, typically zinc, coordinated with organic imidazolate linkers. This specific arrangement gives a rise to an interconnected network of pores and channels, yielding a substantial surface area suitable for gas adsorption and molecular storage[32]. Typically, the surface area of ZIF-8 ranges from several hundred to thousands of square meters per gram (m²/g)[33]. This considerable surface area enables heightened gas adsorption capacity and facilitates efficient surface interactions. Notably, one of the distinctive advantages of ZIF-8 lies in its customizable pore size[34]. By manipulating the choice of metal ions and organic linkers during the synthesis process, the size of the pores can be tailored, thereby allowing customization to suit specific applications and optimizing selective adsorption of desired target molecules[35].

Extending the structure of ZIF-8 with amino- sites presents a promising approach[36–38]. The incorporation of amino-compounds has garnered significant interest in the field of $CO_2$ capture, primarily attributed to their remarkable ability of amine functional groups to chemically interact with $CO_2$ molecules[39]. This chemical interaction referred to as a chemisorption, plays a key role in facilitating the efficient capture and immobilization of $CO_2$ from gas mixtures. Amine sites possess a high affinity for $CO_2$ due to their Lewis basicity[40], allowing for strong bonds to form between the amine functional groups and $CO_2$ molecules[41]. This interaction results in efficient and selective $CO_2$ adsorption, even at low concentrations[13,42]. In summary, the incorporation of amine sites into ZIF-8 presents a promising for enhancing $CO_2$ capture. The affinity, selectivity, regeneration potential and tailoring possibilities offered by amine sites make them critical components in $CO_2$ capture technologies. By selecting different amino-compounds and exploring their structural diversity, it is possible to optimize ZIF-8 for improved $CO_2$ adsorption capacities, selectivity efficiency and stability.

In our approach, we selected 4 different amino-compounds to enhance the $CO_2$ capture efficiency of ZIF-8 based on their distinct structures and length of the hydrocarbon chains. Each amino-compound possesses unique structural characteristics that can affect its interaction with $CO_2$. Tetraethylenepentamine (TEPA) as a polyamine provides a greater number of amine sites due to its multiple amine groups and branched structure[43]. Long-chain alkyl amines, such as

[a.] Centre for Nanomaterials and Biotechnology, Faculty of Science, J. E. Purkyně University, Pasteurova 15, 40096 Ústí nad Labem, Czech Republic
[b.] Department of Solid State Engineering, University of Chemistry and Technology, 166 28 Prague 6, Czech Republic

hexadecylamine (HDA), enhance hydrophobic interactions and potentially increase the adsorption capacity for $CO_2$[44]. On the other hand, ethanolamine (ELA), as a representative of short-chain alkyl amine, contains both primary and secondary amine groups and it offers a combination of chemisorption and physical adsorption mechanisms[45]. Lastly, the cyclic amine structure of cyclopropylamine (CPA) provides unique steric and electronic effects that can affect its affinity for $CO_2$[46]. The behavior of the mentioned amino-compounds and also of their different concentrations into porous ZIF-8 and their effects on $CO_2$ capture was investigated by various characterization methods described in experimental section.

## 2. Experimental

### 2.1. Materials and methods

Zinc nitrate hexahydrate, 2-methylimidazole (2-MIM), tetraethylenepentamine (TEPA), hexadecylamine (HDA), ethanolamine (ELA), cyclopropylamine (CPA) were provided by Sigma-Aldrich, Germany. Methanol (MeOH) and acetone were purchased from Lach-Ner, s.r.o., Czech Republic. All the chemicals were used without further purification.

### 2.2. Synthesis

**Pristine ZIF-8.** The synthesis of ZIF-8 was conducted via a one-pot method at room temperature (RT). To initiate the process, 8 mmol of 2-MIM was dissolved in 50 mL of MeOH, while 4 mmol of zinc nitrate hexahydrate was dissolved in another 50 mL of MeOH. The metal salt solution was carefully added to the ligand solution, resulting in the immediate formation of cloudiness and a white precipitate within the mixture. The solution was thoroughly stirred at 500 rpm for a duration of 24 hours at RT to ensure complete reaction. After the reaction period, the resulting product was separated by centrifugation and subjected to multiple washes using acetone and MeOH to remove impurities. Finally, the ZIF-8 MOF was dried in an oven at 50 °C under atmospheric pressure to obtain the final crystalline material for subsequent characterization and analysis.

**ZIF-8 functionalization.** The procedure for preparing the functionalized ZIF-8 samples remained consistent, with the only variation being the addition of amino-compounds (TEPA, HDA, ELA or CPA) to the ligand solution prior to the introducing the metal salt solution. Three different amounts of amino-compounds (0.5, 1.0 or 2.0 mmol) were used in each sample. Ve kterém kroku byly aminy přidány??

### 2.3. Characterization techniques

**Fourier-transform infrared spectroscopy (FTIR).** The changes in chemistry caused by adding various amino-compounds were sequentially observed. FTIR measurement was performed by Nicolet 6700 spectrometer (Thermo Fisher Scientific, USA) using ZnSe ATR crystal in 600-4000 $cm^{-1}$ region. The FTIR spectra were processed using OMNIC software.

**X-ray Photoelectron Spectroscopy (XPS).** XPS was utilized to investigate the atomic percentages of carbon (C 1s), oxygen (O 1s), nitrogen (N 1s) and zinc (Zn 2p3/2) in surface concentrations. Experimental data were collected using ESCAProbeP spectrometer (Omicron Nanotechnology GmbH, DE). The XPS analysis was performed under ultra-high vacuum conditions using a monochromatic Al K Alpha X-ray source. Subsequently, the acquired XPS spectra were analysed using CasaXPS software to determine the areas of each peak for further evaluation.

**Dynamic light scattering (DLS).** Determination of the hydrodynamic size and size distribution of the nanoparticle colloids, as same as the electrokinetic potential (zeta potential) were evaluated by DLS method. Measurements were performed on Litesizer 500 (Anton Paar, Austria). The instrument is equipped with a 40-mW semiconductor red laser source with a wavelength of 658 nm and the instrument automatically selected the most suitable detection angle (15°, 90° and 175°). The measurements were performed in Omega cuvettes (~300 μl per sample) at RT and samples were analysed immediately after the synthesis. For zeta potential calculation the Smoluchowski approximation method was used.

**Scanning electron microscopy (SEM).** Surface morphology measurements were conducted using a Hitachi SU5000 scanning electron microscope (Hitachi High-Tech Europe GmbH, Germany). All the samples were dispersed in acetone and individually dripped onto a 1×1 cm laboratory glass substrate. After drying, a thin layer of gold was sputtered onto the samples, which were then mounted on standard metal targets using double-sided carbon tape. Measurements were performed under high vacuum using a secondary electron (SE) detector.

**Total specific surface area and pore volume.** Adsorption and desorption nitrogen isotherms were recorded on the Nova3200 instrument (Quantachrome Instruments, USA) and evaluated by using NovaWin software. Firstly, samples were degassed at RT for 24h, then 44-point adsorption and desorption isotherms were measured with the nitrogen (Linde, 99,999% purity) at liquid nitrogen temperature. Total specific surface area was calculated using a 5-point Brunauer-Emmet-Teller (BET) analysis and the Micropore BET Asistant, the pore volume was evaluated by the density functional theory (DFT). Each sample was measured 3× with an experimental error of 5%.

**$CO_2$ capture.** The measurement of the captured $CO_2$ volume was accomplished through $CO_2$ adsorption and desorption isotherms. Prior to analysis, the samples were degassed for 24 hours at room temperature. Subsequently, the adsorption and desorption isotherms were recorded using a gas adsorption instrument Autosorb iQ (Anton Paar, Austria) with $CO_2$ as the adsorbate. All samples were characterized 3× with an experimental error of 5%. The maximum volume of $CO_2$ adsorbed was determined by employing the non-local density functional theory (NLDFT) model provided by the Autosorb software.

## 3. Results and discussion

In the following sections, we will provide a comprehensive discussion on the synthesis and individual characterization techniques employed in this study. The main objective of this characterization is to evaluate their viability and suitability for $CO_2$ capture applications. Our focus will shift towards the successful one-pot synthesis of pristine ZIF-8. Moreover, we will thoroughly investigate the influence of TEPA, HDA, ELA or CPA used for functionalizing ZIF-8. This examination will shed light on the distinct effects of these amino-compounds on the properties and additionally, we will explore the role of varying concentrations of amino-compounds to determine the optimal conditions for achieving desired characteristics in the modified ZIF-8, enabling efficient $CO_2$ capture.

### 3.1. Synthesis of ZIF-8 samples

ZIF-8 MOF was synthesized using a one-pot synthesis approach shown in Fig. 1. The synthesis procedure involved the mixing of a metal salt and a 2-MIM in MeOH as the solvent. Specifically, the metal salt and ligand were combined in a reaction vessel and stirred at RT for a period of 24 hours. During this time, the reaction mixture underwent self-assembly and coordination, leading to the formation of ZIF-8 MOF crystals. The resulting product was then collected, washed, and dried for further characterization and analysis. This one-pot synthesis method offers a convenient and straightforward approach for the preparation of ZIF-8 MOF, allowing for the efficient formation of the desired material.

Furthermore, different amounts of TEPA, HDA, ELA or CPA were added during??? the synthesis of ZIF-8. This approach allowed to systematically study the resulting ZIF-8 samples and to investigate how these changes affect their properties important for $CO_2$ capture. By adjusting the concentrations of these amino-compounds, we sought to elucidate the relationship between the chemical composition of ZIF-8 and its $CO_2$ capture performance.

### 3.2. Characterization of ZIF-8 samples

**Chemical composition.** FTIR and XPS were employed to characterize the surface properties and chemistry of ZIF-8 and its modification with amino-compounds. FTIR analysis provided insights into the chemical structure and functional groups, while XPS allowed for the identification and quantification of surface elemental composition and changes resulting from the amino-compound functionalization.

The FTIR spectra (Fig. 2) illustrate the ZIF-8 modifications introduced by individual amino-compounds at the investigated concentrations. Specifically, the spectra denoted as 2a, 2b, 2c, and 2d correspond to TEPA, HDA, ELA, and CPA, respectively. Additionally, the pristine ZIF-8 spectrum is presented in black for comparison with the modifications.

Pristine ZIF-8 spectrum was analyzed, and the visible peaks were observed in the spectral region spanning 2900-3000 cm$^{-1}$. The first peak, appearing at 3135 cm$^{-1}$, is assigned to C–H stretching vibrations, indicative of the presence of aromatic rings in the ZIF-8 structure. Aromatic moieties are known to contribute significantly to the stability and rigidity of the framework[47]. In addition, a peak at 2930 cm$^{-1}$ is observed, corresponding to C–H stretching vibrations related to the presence of aliphatic groups, which further adds to the complexity and versatility of ZIF-8. A distinctive band at 1584 cm$^{-1}$ is attributed to C=N stretching vibrations, providing evidence of the imidazole rings in the structure, which plays a pivotal role in the formation of the framework, as they act as essential linkers connecting the metal ions. The intense bands observed in the spectral range of 1300-1450 cm$^{-1}$ are associated with the stretching vibrations of imidazole rings[48]. These vibrations arise due to the coordinated nitrogen atoms in the imidazole linkers, which form coordination bonds with metal centers, ensuring the structural integrity. Another significant peak is observed at 1147 cm$^{-1}$, attributed to the stretching vibration of C–N bonds[49]. These C–N bonds within ZIF-8 further confirms the role of imidazole linkers in connecting metal ions and influencing the structural arrangement of the framework. Further insights into the structural arrangement of the C–N bonds and the flexibility of the imidazole ring are provided by the peaks located at 995 cm$^{-1}$, 760 cm$^{-1}$, and 694 cm$^{-1}$. These peaks correspond to C–N bending vibrations and bending vibration of the imidazole ring, respectively[50]. Moreover, the investigation in another study[51] revealed the absence of characteristic features in the FTIR spectrum of ZIF-8 when compared to the spectrum of 2-MIM. Specifically, the disappearance of the broad band spanning the range of 3400-3200 cm$^{-1}$, which is associated with N–H···H hydrogen bond, and the peak at 1843 cm$^{-1}$ corresponding to N–H stretching vibrations. The observed disappearance of these absorption bands indicates that 2-MIM underwent complete deprotonation during the process of ZIF-8 formation. Overall, the analysis of the FTIR spectrum of pristine ZIF-8 reveals distinctive peaks that provide insights into the bonding arrangements and structural features within ZIF-8 framework.

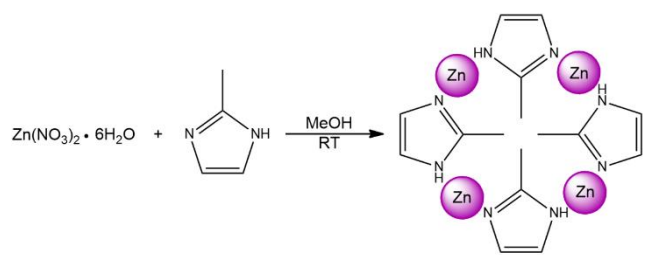

**Fig. 1** Schematic illustration of the ZIF-8 preparation.

FTIR spectra of ZIF-8 with the addition of amino-compounds demonstrate a noticeable increase in the intensity of characteristic peaks that are typical for amine functional groups as the concentration of the amino-compounds increases. This observed trend suggests a direct correlation of the added amines and the increased intensity of the corresponding spectral features, indicative of the successful incorporation and interaction of the amine moieties within the ZIF-8 framework. TEPA (Fig. 1a) is distinguished by a prominent broad band in the spectral range 3000-3600 cm$^{-1}$, two maximums at 3421 cm$^{-1}$ and 3202 cm$^{-1}$, indicating the presence of -N–H and -NH$_2$ stretching vibrations, respectively[52]. Furthermore, spectrum exhibits additional peak in the region 2800-3000 cm$^{-1}$, which can be attributed to CH$_2$ symmetric and asymmetric stretching vibrations. Peak at 1363 cm$^{-1}$, and increased intensity at 1420 cm$^{-1}$ indicate the presence of TEPA-related N–H bending vibrations[52]. Notably, new peak at 1337 cm$^{-1}$ strongly suggests a Zn–N bond to TEPA and its incorporation in the ZIF-8 framework[52,53]. Another distinctive peak emerges at 1100 cm$^{-1}$, which is characteristic of C–N stretching vibrations in the TEPA structure[54]. These peaks further confirm the successful inclusion of TEPA molecules within the ZIF-8 framework, highlighting the formation of coordination bonds between TEPA and metal centers in the ZIF-8. In comparison to TEPA, HDA possesses only one NH$_2$ amino- group in its elongated chain. The spectral analysis reveals a noticeable reduction in the intensity of the broad band observed in the 3000-3600 cm$^{-1}$ region, along with the appearance of a single maximum at 3178 cm$^{-1}$, which corresponds to the NH$_2$ stretching vibration characteristic of HDA[55]. Similar to TEPA, the FTIR spectrum with added HDA exhibits increased intensity of the peak at 1420 cm$^{-1}$ attributed to N–H bending vibrations and additionally, peak at 1335 cm$^{-1}$ belongs to new Zn–N stretching vibrations, thus confirming the effective incorporation of HDA into the ZIF-8. In addition, there is a small peak shift to 1102 cm$^{-1}$, which belongs to the C–N stretching vibrations of HDA. The FTIR spectra with the addition of ELA (Fig. 2c) exhibit considerable resemblance to those acquire with the inclusion of HDA. The similarity in these spectra can be attributed to the chemical interactions and structural changes that occur when these molecules are introduced into the ZIF-7 framework. Both FTIR spectra display comparable vibrational features, such as -NH$_2$ stretching and bending vibrations, which are characteristic of the amino groups present in both ELA and HDA. Both of the amino compounds have linear structure, that allows them to adopt similar conformations when interacting with ZIF-8. This similarity in the orientation of guest molecules can lead to comparable spectral patterns. The FTIR spectra of ZIF-8 with incorporated CPA (Fig. 2d) feature a broad peak in the 3000-3600 cm$^{-1}$ region, with a maximum at 3346 cm$^{-1}$, indicating the presence of NH$_2$ stretching vibrations within the structure. Similar to other spectra analyzed, a distinct peak at 1336 cm$^{-1}$ is attributed to Zn-N bonding. Notably, in comparison to the other amino-compounds investigated, the FTIR spectra of ZIF-8 with added CPA display a prominent peak at 1030 cm$^{-1}$. This peak is characteristic of C–N stretching vibrations specific to the CPA molecular structure[56]. In summary, the FTIR characterization of ZIF-8 with added amino-compounds demonstrates the presence of specific vibrational bands corresponding to N–H, NH$_2$, C–H, CH$_2$, C–N and Zn–N stretching and bending vibrations, signifying the effective incorporation of tested amino-compounds molecules into the ZIF-8 structure.

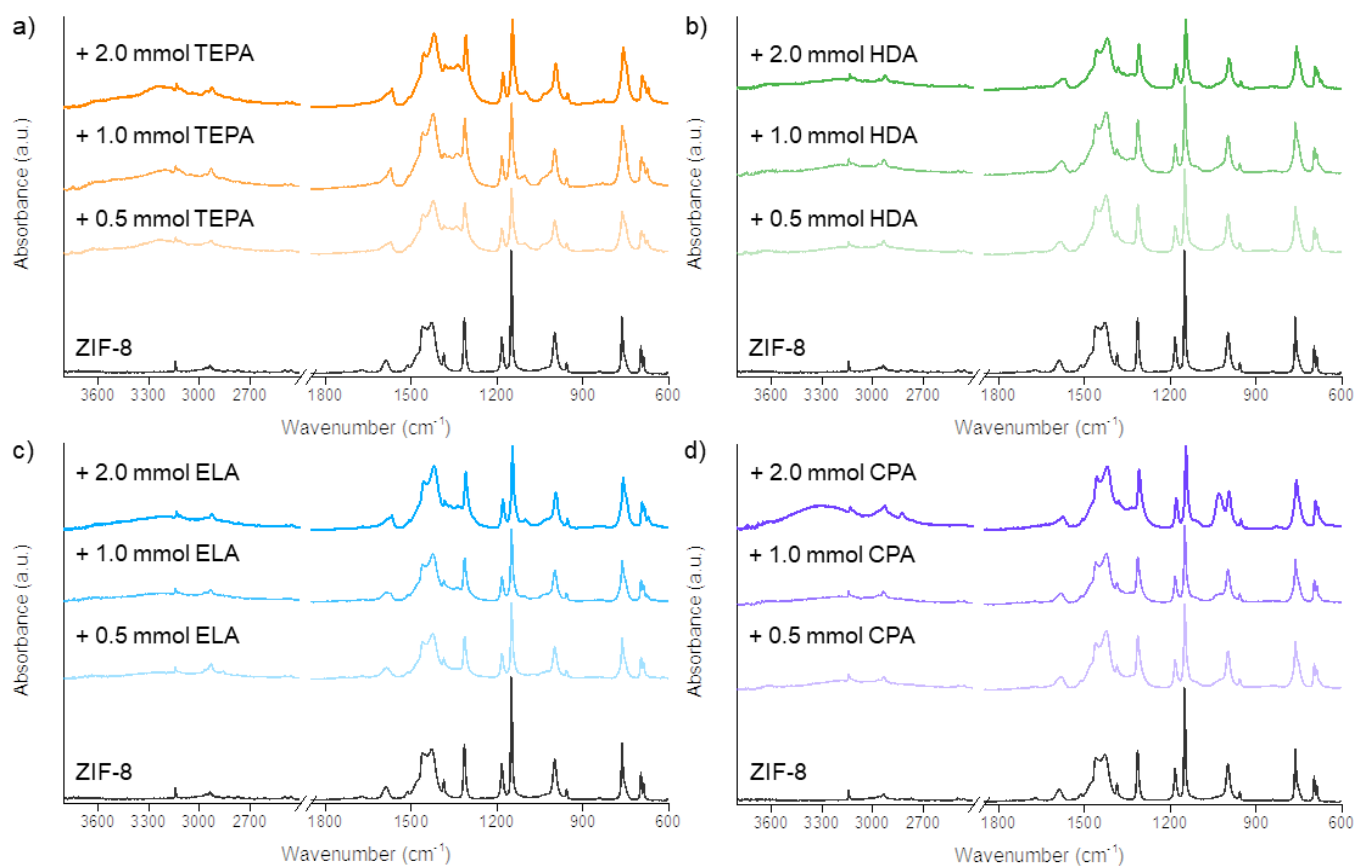

**Fig. 2** FTIR spectra of the pristine ZIF-8 with 0.5, 1.0 or 2.0 mmol a) TEPA, b) HDA, c) ELA or d) CPA.

XPS analysis was conducted to investigate the surface elemental composition of pristine ZIF-8 and its functionalized counterparts with 2 mmol TEPA, HDA, ELA, or CPA. These results of the XPS measurements presented in Table 1 confirmed the presence of C, O, N, and Zn elements on the surface of the pristine ZIF-8, as expected. Following the amine modification, noticeable changes in the percentage ratios between these elements were observed. Among the tested amino-compounds, TEPA expectedly exhibited the highest amino- functional groups content, leading to a 1.6% increase in C and a 3.2% increase in N compared to pristine ZIF-8. On the other hand, functionalization with HDA, which possesses a long hydrocarbon chain with a single amino-group, resulted in a 6.4% increase in C and a 0.9% decrease in N relative to pristine ZIF-8. For ELA, the elemental composition showed a proportional increase of 1.7% in O, attributed to the presence of the -OH group, along with a 0.6% increase in N. Similarly, functionalization with CPA led to a proportional increase of 3.5% in C and 0.8% in N within the structure. It is noteworthy that the addition of compounds containing C and N led to a proportional decrease in the percentage of Zn compared to the other elements. Despite the somewhat similar FTIR spectra of the ZIF-8 samples with added amino-compounds, XPS served as a complementary method, revealing significant changes in their percentages. Particularly, it highlighted the increased presence of C and N in the functionalized structures, indicating the successful integration of the amino-compounds into the ZIF-8 framework.

**Table 1** Percentage of elements in ZIF-8 pristine and with 2 mmol TEPA, HDA, ELA, or CPA added

| Element concentration (at. %) | | | | |
| --- | --- | --- | --- | --- |
| Sample | C (1s) | O (1s) | N (1s) | Zn (2p 3/2) |
| Pristine ZIF-8 | 66.8 | 8.6 | 17.5 | 7.1 |
| 2 mmol TEPA | 68.4 | 6.8 | 20.3 | 4.5 |
| 2 mmol HDA | 72.4 | 6.3 | 16.6 | 4.7 |
| 2 mmol ELA | 66.5 | 10.3 | 18.1 | 5.1 |
| 2 mmol CPA | 70.3 | 6.4 | 18.3 | 5.0 |

**Particle diameter and zeta potential.** Fig. 3 presents the particle size distribution analysis of all samples using DLS. Each subplot in the figures displays the results for pristine ZIF-8 and for the three different amounts (0.5, 1.0, or 2.0 mmol) of each added amino-compound, namely TEPA (Fig. 3a), HDA (Fig. 3b), ELA (Fig. 3c) and CPA (Fig. 3d). It is essential to acknowledge the high sensitivity of the DLS technique to dynamic processes such as aggregation, clustering, and agglomeration, which can significantly impact the observed particle sizes. Moreover, when measuring the hydrodynamic diameter in an aqueous solution, the results include contributions from the solvation layer, consisting of water molecules and ions. Additionally, Fig. 4 presents a graph illustrating the zeta potential values obtained for all samples. The zeta potential is a crucial parameter as it provides valuable insights into the surface charge and stability of colloidal particles. Overall, the combination of DLS data and zeta potential measurements enables a comprehensive characterization of the particle size distribution and surface charge properties of the samples.

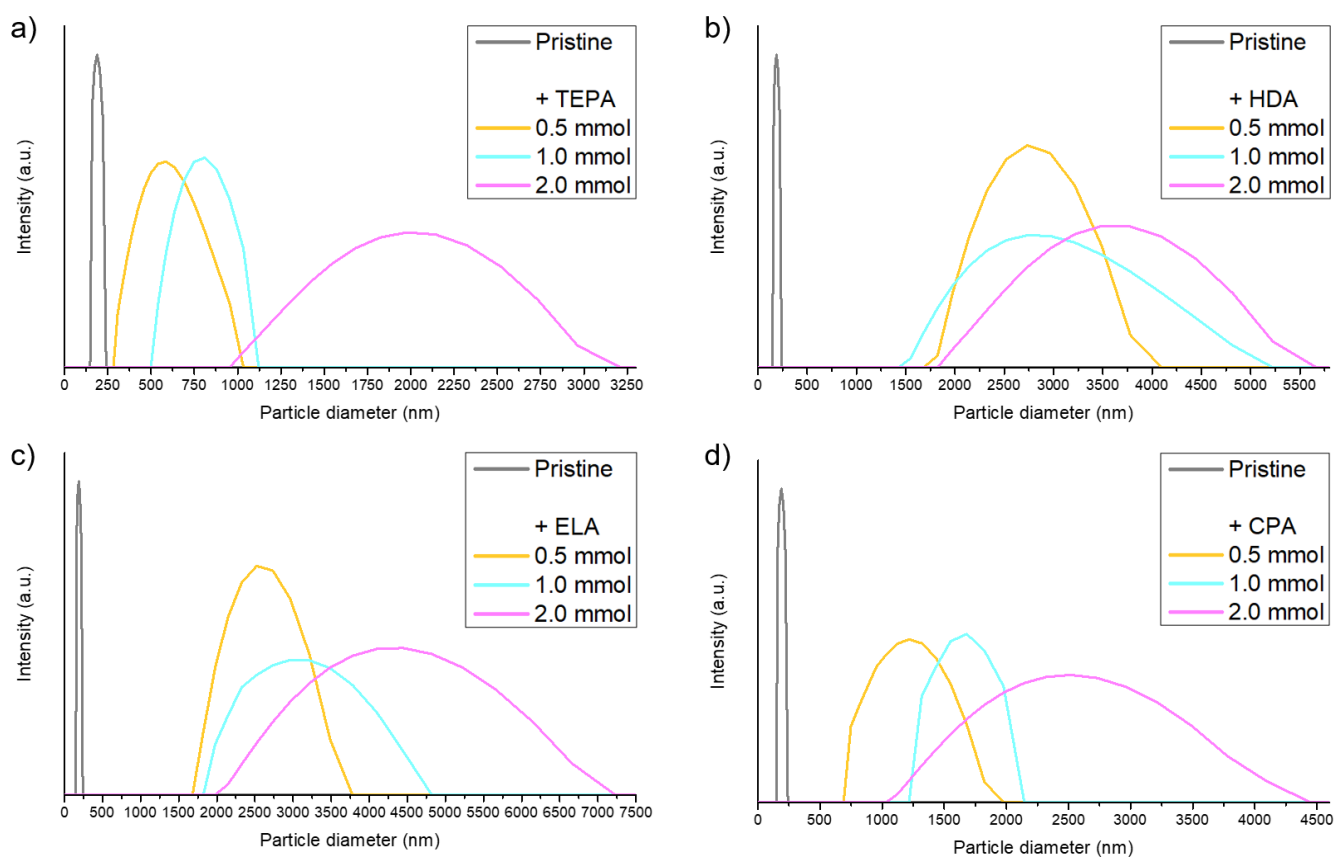

**Fig. 3** Particle diameter distribution of the pristine ZIF-8 and its modification with 0.5, 1.0, or 2.0 mmol a) TEPA, b) HDA, c) ELA or, d) CPA

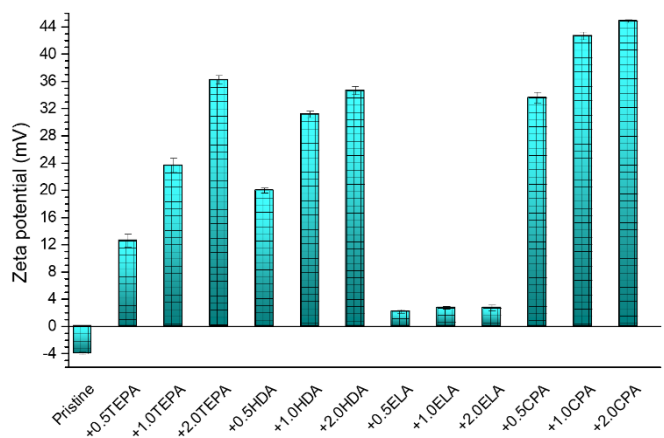

**Fig. 4** Zeta potential values of pristine ZIF-8 and its modification with 0.5, 1.0 or 2.0 mmol TEPA, HDA, ELA or CPA.

The pristine ZIF-8 exhibits a particle size distribution ranging from 145 to 245 nm, with the maximum observed at 188 nm. The polydispersity index (PDI) value of 10.67 indicates that this system has a moderately narrow size distribution. This suggests that the majority of particles are relatively similar in size, although some differences in size exist within the sample. Upon introducing amino-compounds to the ZIF-8 system, we observe a distinct trend across all plots. As the concentration of the amino-compound increases, there is a corresponding increase in particle size variation. This is evident in the broader and more spread-out peaks in the particle size distribution, indicating a polydisperse nature within the samples. The polydispersity signifies considerable variability in particle sizes. Notably, in systems with the highest amino-compound amount (2.0 mmol), the particle sizes become highly heterogeneous, and no clear dominant size is observed. This heterogeneous behavior could be attributed to the interactions and complexities introduced by the higher concentration of amino compounds, leading to a wider range of particle sizes. Overall, the addition of amino compounds to ZIF-8 induces significant changes in the particle size distribution, with the level of variation increasing as the amino-compound concentration rises. The polydispersity observed at higher concentrations underscores the importance of carefully considering the impact of these compounds on the system's particle size uniformity, which is crucial for potential applications.

Pristine ZIF-8 exhibits a negative zeta potential value of -3.97 ± 0.23 mV, indicating a slightly negatively charged surface. However, with the introduction of amino-compounds to the ZIF-8 framework a shift towards positive zeta potential values was observed. These changes are attributed to the presence of functional amino- groups in these compounds, which carry positive charges and interact with the ZIF-8 surface. Notably, the shift towards positive charge values is particularly pronounced for samples containing TEPA, HDA and CPA. Furthermore, as the concentration of these amino-compounds increases, the zeta potential becomes even more positive. This trend indicates that the higher concentration of amino- groups on the surface contributes to a greater positive charge. On the other hand, for ELA, the zeta potential values are also positive but quite similar to each other. This is explained by the fact that ELA contains not only amino- functional groups but also a hydroxyl- group that is negatively charged. The opposing charges of the amino- and -OH groups in ELA likely balance each other out, resulting in zeta potential values that are relatively consistent and positive. In summary, the introduction of amino-compounds to the ZIF-8 structure leads to a shift from the initial negative zeta potential of pristine ZIF-8 towards positive values. The degree of positive charge enhancement is further dependent on increasing amino-compound concentration, except in the case of ELA, where the presence of both of amino- and hydroxyl- functional groups yields relatively consistent positive zeta potential values.

**Morphology.** The morphology of the pristine ZIF-8 samples and its modifications with amino-compounds were investigated using SEM and the micrographs are presented in Fig. 5. The SEM images provide crucial insights into the structural changes induced by the introduction of various amino-compounds, namely TEPA, HDA, ELA and CPA for the 2.0 mmol concentration. The SEM image in Fig. 5a illustrates the characteristic rhombic dodecahedron structure of pristine ZIF-8, which is consistent with the morphology typically observed in one-pot preparation of ZIF-8 using MeOH as the solvent[57,58]. This well-defined structure indicates the successful formation of ZIF-8 crystals under the standard synthesis conditions. Interestingly, upon the addition of TEPA to the ligand solution before introducing the metal salt solution, a noticeable alteration in the ZIF-8 morphology occurs, as shown in Fig. 5b. The distinct rhombic dodecahedron structure is no longer observed, and instead, the SEM image reveals the formation of ZIF-8 clusters with a porous surface. This change in morphology suggests that TEPA influences the nucleation and growth process of ZIF-8, leading to the formation of porous clusters, which may have significant implications for the material's porosity and surface area which will be discussed below. Subsequently, upon introducing HDA or ELA to the synthesis, as depicted in Figs. 5c and 5d, respectively, ZIF-8 clusters exhibit a tendency to aggregate. However, unlike TEPA-induced modifications, no visible pores form on the surface of the clusters in these cases. This suggests that HDA and ELA might influence the crystal growth kinetics differently compared to TEPA. The SEM image of the CPA-modified ZIF-8 sample (Fig. 5e) reveals an interesting phenomenon. The introduction of CPA leads to the formation of 2D flakes between the ZIF-8 clusters. This observation suggests that CPA may interact differently with the ZIF-8 building units, resulting in the generation of unique hybrid structures. These findings open up new possibilities for tailoring the morphology of ZIF-8 through controlled interactions with amino compounds. Comparing all the SEM images, it is evident that the addition of amino-compounds induces changes in the particle size distribution, leading to the formation of a polydisperse system. This phenomenon indicates that the presence of different amino-compounds influences the nucleation and crystal growth processes of ZIF-8, resulting in variations in particle sizes within the sample. It was also confirmed by DLS analyses discussed above.

**Surface area, pore volume and volume of adsorbed $CO_2$**. In Fig. 6, the surface area (Fig. 6a), pore volume (Fig. 6b), and adsorbed $CO_2$ volume (Fig. 6c) results obtained using BET and DFT techniques are presented. The pristine ZIF-8 sample exhibited the highest surface area among all the samples, determined as a value of 1645 ± 9 $m^2$/g, along with the largest pore volume of 0.59 ± 0.02 $cm^3$/g. This indicates that the pristine ZIF-8 has a highly porous structure. It is also visible from SEM results (Fig. 5a). However, the surface area of the amino- modified samples is reduced due to powder aggregation into larger clusters, as observed in the SEM analysis (Fig. 5b-5d). The addition of 0.5 mmol of TEPA causes pore clogging, resulting in significantly surface area. Interestingly, as the TEPA concentration increases, the surface area and pore volume also increase due to the formation of additional pores on the ZIF-8 structure, which is evident from SEM analysis (Fig. 5b). In contrast, the trend is reversed for HDA or CPA modifications. A lower concentration of HDA partially clogs the pores, while a higher concentration leads to significant pore blockage. Similarly, ELA modifications show a trend similar to TEPA, but SEM analysis (Fig. 5e) does not reveal the presence of newly formed pores on the surface.

Regarding $CO_2$ capture, pristine ZIF-8 demonstrates an adsorbed $CO_2$ volume of 31.1 ± 0.4 $cm^3$/g. Notably, some of the modified samples outperform pristine ZIF-8 in terms of $CO_2$ capture efficiency. Specifically, the sample with 2.0 mmol TEPA exhibits 33.3% better efficiency, 0.5 mmol HDA shows 46.6% improvement, and 0.5 mmol CPA displays 18.6% better efficiency than pristine ZIF-8. However, the samples with added ELA do not show significant improvement in $CO_2$ capture compared to pristine ZIF-8. It is worth mentioning that the amount of adsorbed $CO_2$ is influenced not only by the surface area and pore volume but also by the total charge on the surface[59,60], which means the higher presence of amino-functional groups on surface suitable for $CO_2$ interactions. As is shown in Fig. 4, zeta potential measurements reveal that positively charged samples demonstrate the better efficiency in $CO_2$ capture. This suggests that the electrostatic interaction between the adsorbent surface and $CO_2$ molecules plays a significant role in the capture process. The modification of ZIF-8 with different amino-compounds affects its surface area, pore volume and ultimately the efficiency of $CO_2$ capture. The results demonstrate that samples with optimal modifications, such as 2.0 mmol TEPA or 0.5 mmol HDA, exhibit significantly improved $CO_2$ capture efficiency. These findings shed light on the potential of suitable amino-compound-modified ZIF-8 materials for $CO_2$ capture applications and highlight the importance of surface charge in influencing the adsorption process.

## 4. Conclusions

In this study, we have successfully synthesized and characterized amino-compounds-modified ZIF-8 materials for $CO_2$ capture applications. The one-pot synthesis method provided the convenient and straightforward approach for the preparation of ZIF-8 MOF with high efficiency for CO2 capturing. By systematically varying the concentrations of TEPA, HDA, ELA or CPA, we investigated the distinct effects of these studied amino-compounds on the properties of ZIF-8, important for $CO_2$ capture. FTIR analysis revealed the successful incorporation of the amino-compounds into the ZIF-8 framework, as evidenced by the characteristic peaks corresponding to N-H, $NH_2$, C-H, $CH_2$, C-N, and Zn-N stretching and bending vibrations. XPS measurements further confirmed the integration of the functional amino- groups into the ZIF-8 structure, leading to changes in the elemental composition of the material. DLS and zeta potential measurements provided valuable insights into the particle size distribution and surface charge properties of the modified ZIF-8 samples. The introduction of amino-compounds led to a polydisperse system with increased positive zeta potential values, except for ELA, which exhibited relatively consistent positive values due to the presence of both amino- and hydroxyl- functional groups. SEM analysis revealed the morphological changes induced by the amino-compounds, resulting in the formation of porous clusters, aggregated particles and unique hybrid structures. The modification process influenced the nucleation and crystal growth of ZIF-8, leading to variations in particle sizes within the samples. The surface area and porosity analysis showed that the surface area and pore volume of the modified samples were reduced compared to pristine ZIF-8 due to powder aggregation into larger clusters. However, some of the modified samples displayed improved $CO_2$ capture efficiency compared to the pristine ZIF-8, with the optimal modifications of 2.0 mmol TEPA, 0.5 mmol HDA and 0.5 mmol CPA showing significantly better performance. Overall, this comprehensive study on selected amino-compound-modified ZIF-8 materials provides valuable insights into their potential for $CO_2$ capture applications. The results highlight the importance of surface charge and morphology in influencing the adsorption process, paving the way for the design and optimization of novel materials for efficient $CO_2$ capture and mitigation of greenhouse gas emissions. Further research in this direction could contribute to addressing the urgent global challenge of climate change and promote the development of sustainable energy technologies.

## Author Contributions

V. Neubertová: Conceptualization, Methodology, Validation, Investigation, Writing – original draft, Visualization. V. Švorčík: Validation, Resources, Visualization. Z. Kolská: Validation, Resources, Supervision, Project administration.

## Conflicts of interest

There are no conflicts to declare.

## Acknowledgements

This work was supported by the Czech Science Foundation GAČR under project no. 23-05197S and by the Grant Agency J. E. Purkyně University in Ústí nad Labem under project No. UJEP-SGS-2022-53-004-2 and This work was supported by the Project OP JAK_Mebiosys, No.

CZ.02.01.01/00/22_008/0004634 of the Ministry of Education, Youth and Sports, which is co-funded by the European Union.
Data Availability Statement
The data presented in this study are available at https://doi.org/10.5281/zenodo.10392555.